# Drive the Dirac Electrons into Cooper Pairs in Sr$_x$Bi$_2$Se$_3$

Guan Du[1], Jifeng Shao[2], Xiong Yang[1], Zengyi Du[1], Delong Fang[1], Changjing Zhang[2,3†], Jinghui Wang[1], Kejing Ran[1], Jinsheng Wen[1,3], Huan Yang[1,3†], Yuheng Zhang[2,3] and Hai-Hu Wen[1,3†]

[1] National Laboratory of Solid State Microstructures and Department of Physics, Nanjing University, Nanjing 210093, China

[2] High Magnetic Field Laboratory, Chinese Academy of Sciences and University of Science and Technology of China, Hefei 230026, China

[3] Collaborative Innovation Center of Advanced Microstructures, Nanjing University, Nanjing, 210093, China

**Topological superconductor is a very interesting and frontier topic in condensed matter physics[1]. Despite the tremendous efforts in exploring the topological superconductivity, its presence is however still under heavy debates[2-11]. The Dirac electrons are supposed to exist in a thin layer of the surface of a topological insulator[12]. Due to the finite spin-orbital coupling, these electrons will have a spin-momentum locking effect. In this case, the superfluid with the spin singlet Cooper pairing is not completely comforted by the Dirac electrons[13]. It thus remains unclear whether and how the Dirac electrons fall into Cooper pairing in an intrinsic superconductor with the topological surface**

**states[3]. In this work, we show the systematic study of scanning tunneling microscope/spectroscopy on the possible topological superconductor $Sr_xBi_2Se_3$. We first show that only the intercalated (or inserted), not the substituted Sr atoms can induce superconductivity. Then we show the full superconducting gaps without any abnormal in-gap density of states as expected theoretically for the bulk topological superconductivity[7]. However, we find that the surface Dirac electrons will simultaneously condense into the superconducting state when the energy is smaller than the bulk superconducting gap. This vividly demonstrates how the surface Dirac electrons are driven into Cooper pairs.**

After the demonstration of topological insulators, the search for topological superconductors has already become a hot topic in condensed matter physics, in one dimensional[14,15], two dimensional (2D)[16,17], and three dimensional (3D) systems[4-11]. Theoretical criteria for defining a 3D topological superconductor (TSC) has been proposed[6] and several materials were deemed to be the candidates of 3D TSC, such as doping induced superconductors from topological insulators[6,7] and topological crystalline insulators[10]. Point contact measurements have detected zero-bias conductance peaks which are interpreted to be signatures of Majorana fermions[4,5,10] of TSC. However, the STM studies of $Cu_xBi_2Se_3$[9] and $(Pb_{0.5}Sn_{0.5})_{0.7}In_{0.3}Te$[11] give opposite views about the presence of topological superconductivity in these systems. Alternatively, it is shown that the Dirac electrons on the surface may be driven into Cooper pairs by the superconducting proximity effect[13]. 2D topological superconductivity will arise and a Majorana Fermion will emerge at the vortex core according to theoretical proposals[13] and this seems getting the support from the experiments of hetero-structures[16,17]. However, a vivid demonstration of driving the surface Dirac electrons into the Cooper pairs is still lacking.

Recently, superconductivity has been discovered in $Sr_xBi_2Se_3$ which is supposed to be a promising candidate of TSC[18-21]. Experimentally the quantum oscillation with the Dirac dispersion has been detected in the measurement of global resistivity in the normal state[18], which is supported by the data of angle resolved photo-emission (ARPES)[20]. However, it is unclear whether and how these surface topological electrons will condense into the Cooper pairs. Nevertheless, the existence of topological surface states, lower Fermi level, better showed bulk superconductivity comparing to $Cu_xBi_2Se_3$ make $Sr_xBi_2Se_3$ a good platform to study the possible topological superconductivity[20].

We have grown the superconducting $Sr_xBi_2Se_3$ crystals using flux method[18] and conducted systematic studies using scanning tunneling microscope/spectroscopy (STM/STS). The samples generally exhibit sharp superconducting transitions which are comparable with the previous reports[18,19,21], and the characterization of the basic properties is presented in Supplementary Information I. Our samples, like $Cu_xBi_2Se_3$, display clear inhomogeneity in general. In our STM/STS studies, two kinds of regions with significant distinctions are found. The first kind has a very clean and atomically flat surface as shown in Fig.1a. The impurities can be viewed as substituted Sr atoms and show up as triangular like images, as viewed by the atomically resolved topography in Fig.1a and 1c. As presented in Fig.1b, the STS obtained in this region does not possess an evident superconducting gap at low temperatures. The local height of the impurity is only slightly enhanced (Fig.1a) and the STS without superconducting gapped feature is quite general in this region, and varies little when traverse crossing the substituted Sr, as shown by the bottom panel of Fig.1c. The second kind of region has terraces decorated by clusters with much larger height scattering when crossing them, as shown in the middle of Fig.1d. Those clusters vary in size and height and mostly are movable by the STM tip. We presume that the clusters are composed by intercalated or inserted Sr atoms, just like the Cu clusters in $Cu_xBi_2Se_3$[2,9]. Concerning the difficulty of intercalating Sr atoms with much larger radii into the van der Waals spacing, it may be possible that Sr clusters are formed in the growing process with fault growth along c-axis, and the cleaving is easier to occur at this layer. Interestingly, in a region with many Sr clusters, the STS exhibits a clear superconducting gap (see Fig.1e). Substituted Sr atoms also exist in the superconducting region and turn out to have little effect on superconducting gap (Fig.1f). We assume that the Sr substitution, probably doped to the Bi sites, behaves as an hole doping concerning the different cationic states of $Sr^{2+}/Bi^{3+}$, while the Sr clusters remaining in the van der Waals gap act as electron doping. If this scenario is correct, it means that electron doping by the intercalated Sr can induce superconductivity in $Bi_2Se_3$, as in the case of $Cu_xBi_2Se_3$.

According to the theoretical proposal for TSCs[6], $Sr_xBi_2Se_3$ is considered to be a promising candidate. To verify its topological property, analyzing the superconducting pairing symmetry is very essential. In the region with well-formed superconductivity, we manage in detecting the STS with the bottom near zero-bias energy almost touching zero. The Sr clusters in big size can be viewed everywhere in the region (Fig.2a). The blurry transverse lines in Fig.2a are caused by dragging the clusters during scanning, demonstrating the mobility of the Sr clusters. When going across in this region, as highlighted by the white arrowed line in Fig.2a, the shape of the STS and the gap values are quite homogeneous according to the spectra measured at 400 mK as shown in Fig.2b. The spectrum displayed in Fig.2c is the average of all the data in Fig.2b. The STS with pronounced coherence peaks and perfect suppression of *dI/dV* within the gap look much better than the previous STM/STS results in $Sr_xBi_2Se_3$[20]. The "U" shape instead of a "V" shape feature suggests that the superconducting gap is nodeless. The density of states (DOS) within the superconducting gap is fully gapped and the remaining finite zero-bias DOS is due to a slight scattering/broadening effect[22]. There is no abnormal DOS inside the gap range which contradicts the theoretical proposals of a 3D TSC[7]. This may suggest that the superconductivity in bulk Sr$_x$Bi$_2$Se$_3$ is topologically trivial.

To have a further comprehension on the data, we fit the STS with different gap functions. As shown in Fig.2c-d, the spectrum can be nicely fitted by the Dynes model[22] using two gaps, one with an s wave and another with a slightly anisotropic s wave gap, with $\Delta_s$ = 1.15 meV and $\Delta_{ani-s} = 1.37(0.19\cos 4\theta + 0.81)$ (meV). As a comparison, we fitted the data with several other scenarios of different superconducting gaps, as shown in Fig.2d, verifying that the double gap fitting with s + ani-s wave can interpret the nature of the superconductivity in the material very well. We have also conducted studies by adjusting the fraction of two gaps in s + ani-s wave fitting, and a combination of 26%$\Delta_s$ + 74%$\Delta_{ani-s}$ turns out to be the best fit (Fig.2c-d). The detailed fitting results are given in Supplementary Information II (see also Fig.S2). We hence suggest that there should be two bands with different gaps existing at the Fermi energy, and this consists with the recent ARPES study of Sr$_x$Bi$_2$Se$_3$ that the topological surface state coexists with the bulk state around the Fermi energy[20]. According to the ARPES data, the Fermi surfaces of Sr$_x$Bi$_2$Se$_3$ are two centric circles with the same size around $\bar{\Gamma}$ point (with probably a slight hexagonal distortion)[3]. Because of the spin non-degeneracy, the

DOS of the topological surface state on Fermi surface is about half of that in bulk state. Taking the z-axis dispersion of the bulk state into account, the proportion of the DOS of the two components on the Fermi surface between topological surface electrons and normal bulk electrons should be less than 33% and more than 67%, respectively. From above analysis, we can conclude that the bigger gap with the ratio of 74% maybe contributed by bulk superconductivity and the smaller one with the ratio of 26% is due to the Dirac electrons of the topological surface state in the superconducting state. Full gapping of STS manifests itself that the Dirac electrons of topological surface state have been driven into Cooper pairs.

As displayed in Fig.3a, we further measured tunneling spectra at different temperatures. The superconducting feature vanishes at about $T_c \approx 5$ K, which is higher than that in the transport and magnetization measurements (Fig.S1a). The STS at different temperatures can be well fitted by the s + ani-s wave gaps with the fixed proportion as quoted above. As shown in Fig.3b, the temperature dependent gap values obey the BCS model. Fig.3c shows the magnetic field dependence of tunneling spectra verifying that the $H_{c2}$ can be as high as 5 T which is also larger than that in the transport measurements (Fig.S1d). Note the STS shown here at different magnetic fields is obtained by averaging many STS measured at different locations in the same area under the same field, therefore the oscillation of *dI/dV* due to the Landau level (LL) effect is not seen here.

With a magnetic field of 5 T, as shown in Fig.3d, we succeed in observing the LL peaks near the Fermi energy in a clean region over the length scale of about 300 nm with many surrounding Sr clusters. This STS is obtained at a fixed location. As has been studied in plenty of works, the LLs of 2D surface state can be easily viewed by STS and the 3D bulk state LLs are absent[23-27]. The enhancement of the LL peaks near Fermi energy is due to the much enhanced quasiparticle lifetime in approaching the Fermi energy[23]. The strong LL peaks persist in superconducting $Sr_xBi_2Se_3$ indicating that the surface Dirac electrons stay intact and do not merge into the bulk band at Fermi level. This along with the double gap fitting referred above makes it plausible that the surface Dirac electrons have been driven into superconductivity, which may be caused by the proximity effect of the bulk superconductivity.

In order to vividly show whether and how the Dirac electrons are driven into Cooper pairs in $Sr_xBi_2Se_3$, we did a systematic research on the LL peaks in the area mentioned above. Since the Fermi energy is within the valence band of the bulk state, the Dirac point is quite far away from the Fermi level (about 340 meV, see Supplementary Information III). Therefore, any tiny spatially local alteration to the electronic state, like the in-plane stress which might mildly influence the surface state dispersion, will shift the LL peak positions[26]. The surface of this area has a slight fluctuation which can be caused by the intercalated Sr beneath the surface layer leading to slight stress in plane, or the inhomogeneous doping level of electrons by Sr clusters. Therefore, the random behavior of LL peaks over a large region can be viewed and understandable. This can offer us the opportunity to fetch the information of Dirac electrons by observing the amplitude of the LL fluctuations if the spectra measured at a fixed magnetic field are plotted together. We measured the STS at magnetic fields of 0, 2, 3, 4, 5 T in the same area, the normalized data are shown in Fig.4a to 4e. The data treating and normalizing process are presented in Supplementary Information IV. The enhancement of LL fluctuations and the suppression of superconductivity versus magnetic field can be seen clearly here. One can find that the STS are rather homogeneous at 0 T, as presented in Fig.4a. However, at 2T, a waving like signal of *dI/dV* outside the superconducting gap region shows up, and gets stronger with a higher magnetic field. These waving like signal is actually due to the LLs with a spatial fluctuation effect[26]. In order to see it more clearly, we take an average of all STS measured at the same magnetic field but different positions, and then overlay the data together with their averaged curve as shown by the red solid lines in Fig.4f to 4j. One can clearly see that the spectra outside the superconducting gap fluctuate around the average value and the fluctuating magnitude increases with the magnetic field, indicating that the fluctuation is caused by the LLs rather than the intrinsic noise and background. We then subtract the average spectrum from the data measured at different locations, leaving the fluctuation only and plot them together for each field, as shown in Fig.4k-4o. At zero field, as shown in Fig.4k, the fluctuating band width is quite narrow. Here the slight difference of the band width between inside and outside of the superconducting gap region is caused by the finite background floating of the spectra obtained at different positions. When the magnetic field is 2T the fluctuation band width outside the superconducting gap is getting much stronger, but the band width within the gap rapidly shrinks to a smaller value (Fig.4l). At a magnetic field of 3T, the superconducting gap is still visible (Fig.4h) and the fluctuating band width is getting stronger compared with 2T. Under the magnetic fields of 4T and 5T, the gapped feature is hardly visible in the raw and overlaid data, which means that most of the Cooper pairs are broken. It is reasonable to take the amplitude of the fluctuating band width from Fig.4l to 4o as the index of the population of the Dirac electrons. If the Dirac electrons stay unpaired, the fluctuation band width due to the LLs would be the same within and outside the gap. The shrinking of the fluctuating band width of LLs within the gap at 2T (probably also seen partly at 3T) gives the smoking gun evidence of driving the Dirac electrons of the surface topological state into Cooper pairs and they simultaneously condense into the superconducting state.

For the 3D TSC, it is predicted that the surface Andreev bound state appears within the bulk quasiparticle gap giving rise to significant zero-bias DOS on the surface and making the coherence peaks suppressed or absent[7,8]. This contrasts sharply with our STM/STS results, indicating that $Sr_xBi_2Se_3$ may not a bulk TSC. From our theoretical fitting to the spectra in Fig.2c, it can be concluded that the bulk component of the superconducting gap is slightly anisotropic. Recently, Matano *et al.* [28] observed a two-fold angle dependence of the Knight shift in the NMR measurements in $Cu_xBi_2Se_3$, which gets a qualitative support from the specific-heat measurements by Yonezawa *et al.*[29] This may be explained as the possible nematic superconducting state[30] or the triplet pairing in this material. However, our experiment proves that the topological surface state persists independently from the bulk band at the Fermi level and the Dirac electrons are driven into Cooper pairs at low temperatures. We must emphasize that the induced Cooper pairs constructed by the Dirac electrons of the surface state may behave differently from the Cooper pairs in the bulk, since the former has a spin-momentum locking effect. In this case, a partial spin triplet component may be expectable[13]. This interesting expectation calls for future efforts. We can reasonably assume that the condensed Dirac electrons of the surface state discovered in this work may form a 2D TSC, which will spontaneously realize a Majorana mode in the vortex core that exhibit non-Abelian quantum statistics[13]. Our experiment performed in $Sr_xBi_2Se_3$ certainly makes a step further in helping to search the TSC and the Majorana Fermions in an intrinsic superconducting system.

**Methods**

$Sr_xBi_2Se_3$ single crystals were grown by flux method. For STM/STS measurements we cleaved the samples in an ultrahigh-vacuum chamber with a base pressure of about $10^{-10}$ Torr. The STM studies were carried out with an ultrahigh-vacuum, low temperature and high magnetic field scanning probe microscope USM-1300 (Unisoku Co., Ltd.) with pressure better than $10^{-10}$ Torr. The STS spectra were measured by a lock-in amplifier with an AC modulation of 0.1 mV at 987.5 Hz to lower down the noise.

**Acknowledgements** We thank Yong Chen at Purdue University, Qianghua Wang at Nanjing university, Genda Gu at Brookhaven National lab for helpful discussions. This work was supported by national natural science foundation China (NSFC) with the projects: A0402/11534005, A0402/11190023, A0402/11374143, A0805/U1532267; the Ministry of Science and Technology of China (973 projects: 2012CB821403) and PAPD.

**Author Contributions**

The low-temperature STS measurements were performed by G.D., X.Y., Z.Y.D., D.L.F, H.Y., and H-H.W. The samples were prepared by J.F.S, C.J.Z, J.H.W., J.S.W, K.J.R and Y.H.Z. The simulation based on the Dynes model and the data analysis was performed by G.D. and X.Y. H-H.W. coordinated the whole work and wrote the manuscript with G.D.. All authors have discussed the results and the interpretations.

**Author Information** The authors declare no competing financial interests. Correspondence and requests for materials should be addressed to H-H.W (hhwen@nju.edu.cn), or C-J. Z. (zhangcj@hmfl.ac.cn), or H. Y. (huanyang@nju.edu.cn).

Figures and legends

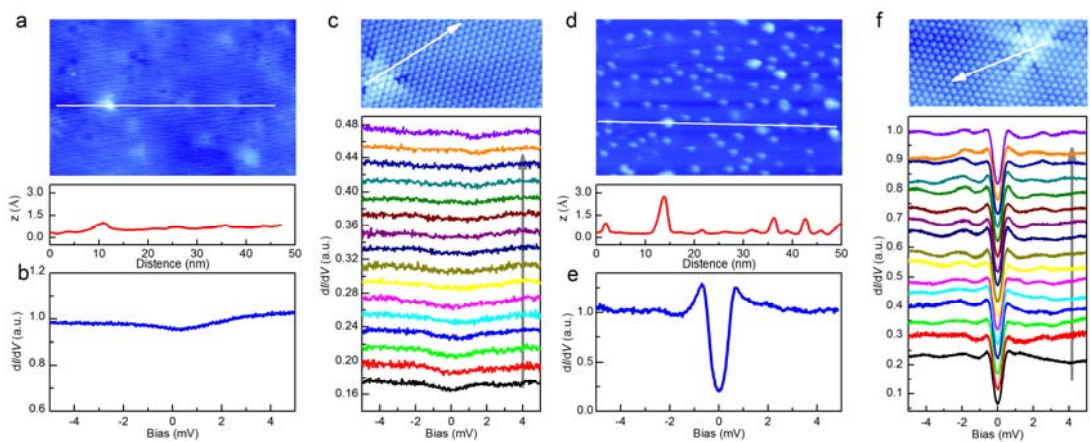

**Figure 1 | Topographic STM images and tunneling spectra. a,d**, Topographic images of the non-superconducting (superconducting) region (50 nm × 32 nm) with different density of the Sr clusters. The center of a triangular image in **a** is roughly corresponding to a substituted Sr impurity. Substituted Sr also exists in the region with well-formed superconductivity with more Sr clusters. The curve below **a** and **d** show the height distribution measured along the white lines marked in **a** and **d**. **b,e**, A typical tunneling spectrum measured at 400 mK in non-superconducting (superconducting) area with the corresponding topographies shown in **a** and **d**. **c,f** shows the atomic resolved topography of non-superconducting (superconducting) regions in a field of view of 10nm×5nm. The spatially resolved tunneling spectra taken at zero field and along the arrowed lines are shown below the images.

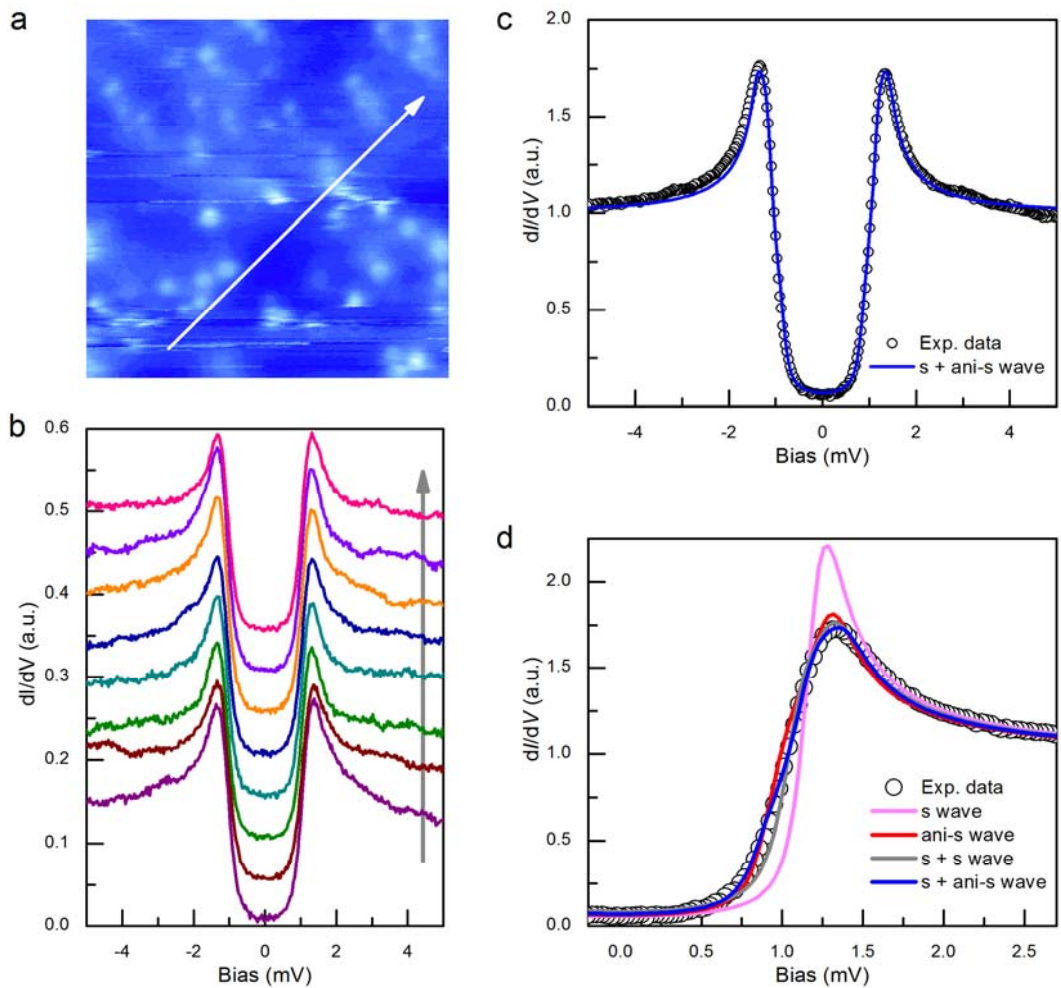

**Figure 2 | Full superconducting gaps and theoretical fittings. a**, The topography of the superconducting area (50 nm × 50 nm) with quite dense Sr clusters. **b**, The spatially resolved tunneling spectra taken along the arrowed line in **a** at 400mK. **c**, The averaged and normalized spectrum of the spectra in **b**. The solid curve is the theoretical fitting with Dynes model using s + ani-s wave. **d**, The comparison of theoretical fittings with several different superconducting gap functions. The fitting parameters are presented in Supplementary information II.

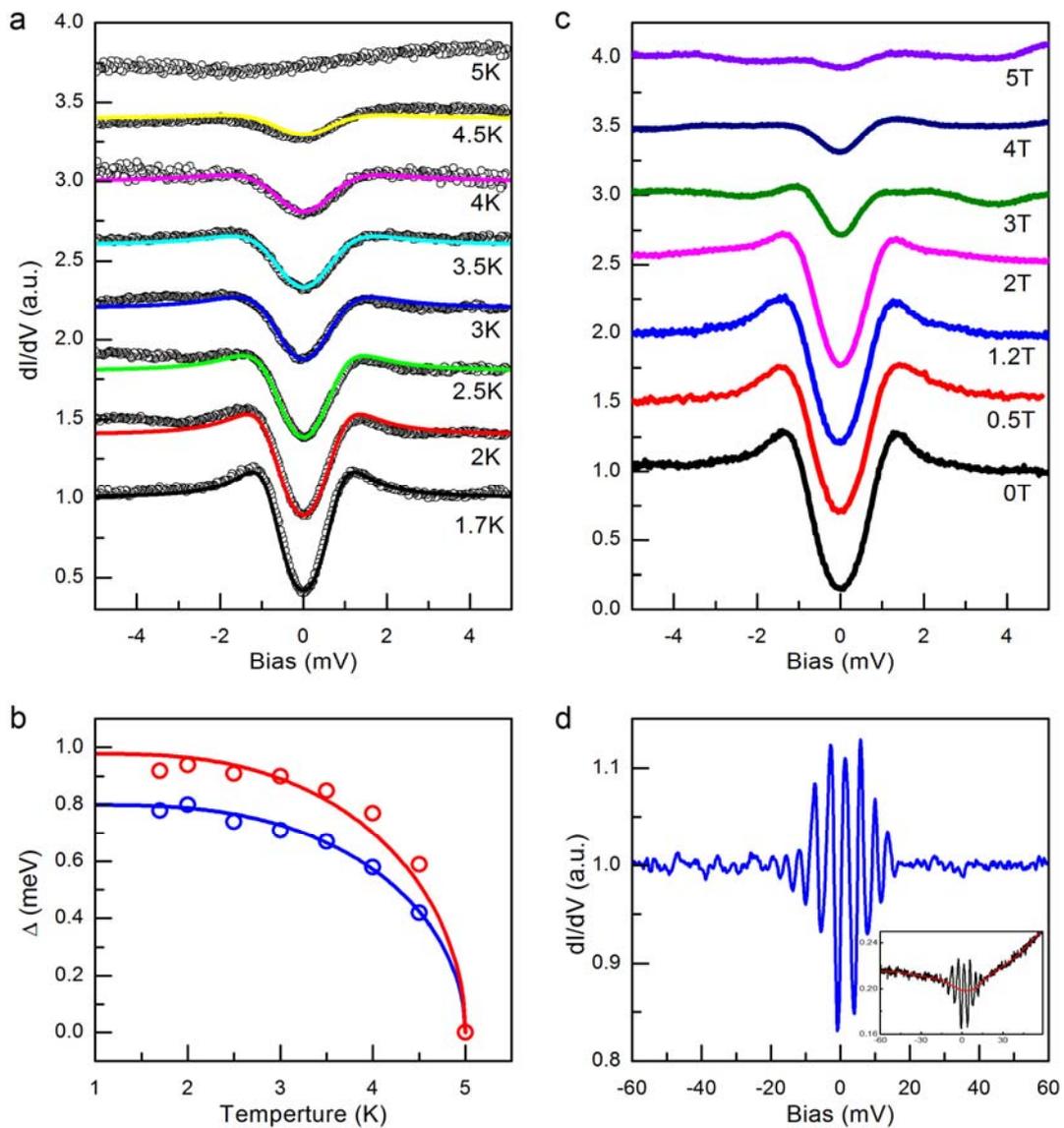

**Figure 3 | Temperature and magnetic field dependence of tunneling spectra, and the Landau levels. a**, The evolution of the STS spectra with temperature increased from 1.7 K to 5 K at zero field. The superconducting gaped feature vanishes gradually at about 5K. The solid lines are theoretical fittings using s + ani-s wave. **b**, The temperature dependence of $\Delta_s$ (blue dots) and $\Delta_{ani-s}$ (red dots) obtained by fitting. The solid lines are obtained through the numerical solution to the BCS gap equation by

fixing $\Delta_s(0)$ = 0.8 meV, $\Delta_{ani-s}(0)$ = 0.98 meV and $T_c$ = 5 K derived from **a**. The gap values are different from that in Fig.2 since the measurements are made at different positions with different local conditions. **c**, Evolution of the STS spectra with magnetic field increased from 0 to 5 T. The $H_{c2}$ is around 5 T. Here for each field, the STS is obtained by averaging the STS data measured at different locations, therefore the oscillations due to the LLs is not visible. **d**, The *dI/dV* spectrum measured at 5 T at a fixed location (smoothed by averaging about 20 STS curves measured at this location) with the background subtracted. The spectrum has been smoothed by averaging the neighboring 10 data points to lower down the noise. The black curve of inset is the raw data. The background is obtained by averaging the neighboring 200 data points shown as the red curve in the insert.

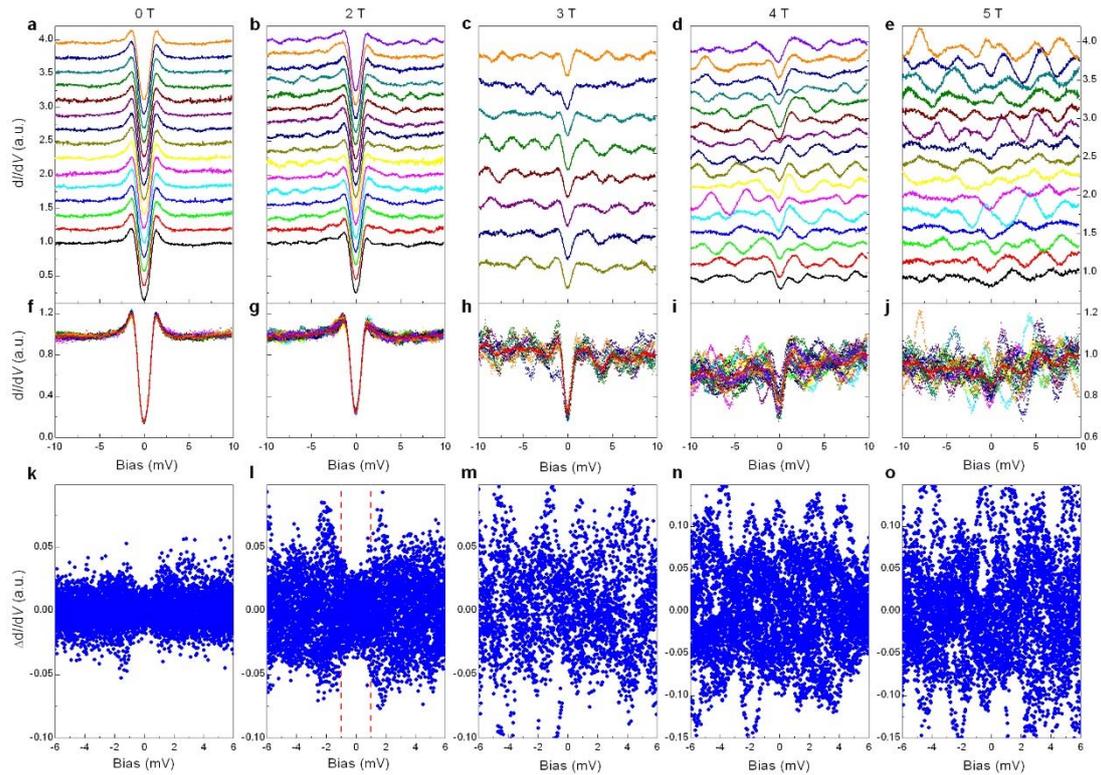

**Figure 4 | The variation of LL fluctuations with the increase of magnetic field. a** to **e,** The spatially resolved spectra obtained over the same area with the magnetic field of 0, 2, 3, 4, and 5 T, respectively. **f** to **j,** The experimental data plotted together with the average spectrum at each magnetic field. The red curves are the average spectra. The way to normalize the spectra is presented in Supplementary Information IV. **k** to **o**, The difference between the normalized spectrum at different locations and the averaged curve, this exemplifies only the fluctuation due to the LLs. The dashed vertical lines in l show the approximate gap value at 2T. One can see that the fluctuation due to the LLs is strongly suppressed within the gap, which can be seen from both **g** and **l**.

# SUPPLEMENTARY INFORMATION

**Supplementary Information I**

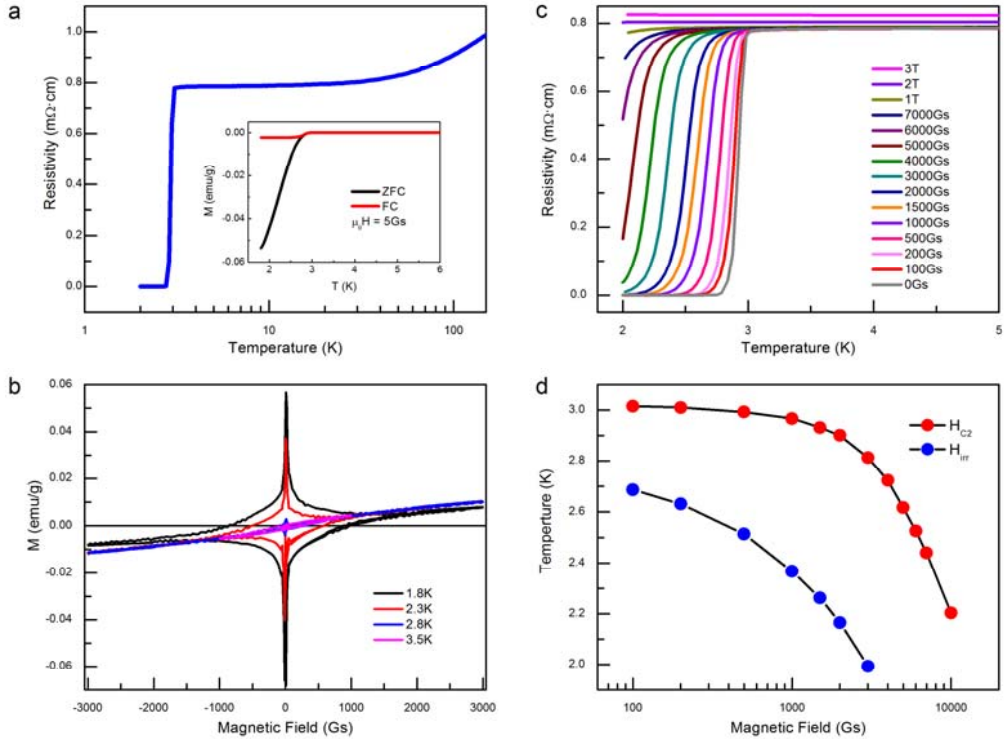

**Figure S1 | Physical characterization of the $Sr_xBi_2Se_3$ single crystal. a**, Temperature dependence of resistivity of a $Sr_xBi_2Se_3$ single crystal at zero field. The inset shows the temperature dependence of magnetic susceptibility measured with zero-field-cooled (ZFC) and field-cooled (FC) processes at an applied magnetic field of 5 Oe. the superconducting transition temperature is about 3 K. **b**, The M-H loop as a function of temperature with the magnetic field applied parallel to *c*-axis. **c**, The resistivity as a function of magnetic field applied parallel to *c*-axis. **d**, Phase diagram of the irreversibility magnetic field ($H_{irr}$) and $H_{c2}$ obtained by 1%

and 99% of the normal state resistivity $\rho_n$. the value of $H_{c2}$ is deduced to be about 3.5 T.

**Supplementary Information II**

The spectrum in Fig.2c was fitted with tunneling current for one gap[1] of

$$I(V) \propto \int_{-\infty}^{\infty} d\varepsilon \int_{0}^{2\pi} d\theta [f(\varepsilon) - f(\varepsilon + eV)]$$

$$\times \text{Re}\left\{\frac{\varepsilon + eV - i\Gamma}{[(\varepsilon + eV - i\Gamma)^2 - \Delta^2(\theta)]^{1/2}}\right\}$$

where $\Gamma$ is the broadening parameter, $f(\varepsilon)$ is the Fermi distribution function which contains the thermal broadening effect at some finite temperatures, and the temperature parameter used for fitting is the same as the experimental temperature. The fitting results with different gap functions are shown in Fig.2d. The pure single s-wave fitting to the spectrum yields a superconducting gap value of $\Delta$ = 1.2 meV and broadening parameter of $\Gamma$ = 0.07 meV. For a single anisotropic s-wave gap, the fitting leads to $\Delta$ = 1.3 meV and broadening parameter of $\Gamma$ = 0.07 meV. For the double s-wave fitting, the two mixed components have the proportion of 40%$\Delta_1$ + 60%$\Delta_2$ with $\Delta_1$ = 1.3 meV, $\Gamma_1$ = 0.09 meV, $\Delta_2$ = 1.05 meV, $\Gamma_2$ = 0.09 meV. For the s + ani-s wave fitting, the two components have the proportion of 26%$\Delta_s$ + 74%$\Delta_{ani-s}$ with $\Delta_s$ = 1.15 meV, $\Gamma_s$ = 0.075 meV, $\Delta_{ani-s}$ = 1.37(0.19 cos 4θ + 0.81) meV ,

$\Gamma_{ani-s}$ = 0.075 meV. The proportion of 26%$\Delta_s$ + 74%$\Delta_{ani-s}$ is obtained through searching the minimum of the lowest root-mean-square (RMS), as shown in Fig.S2.

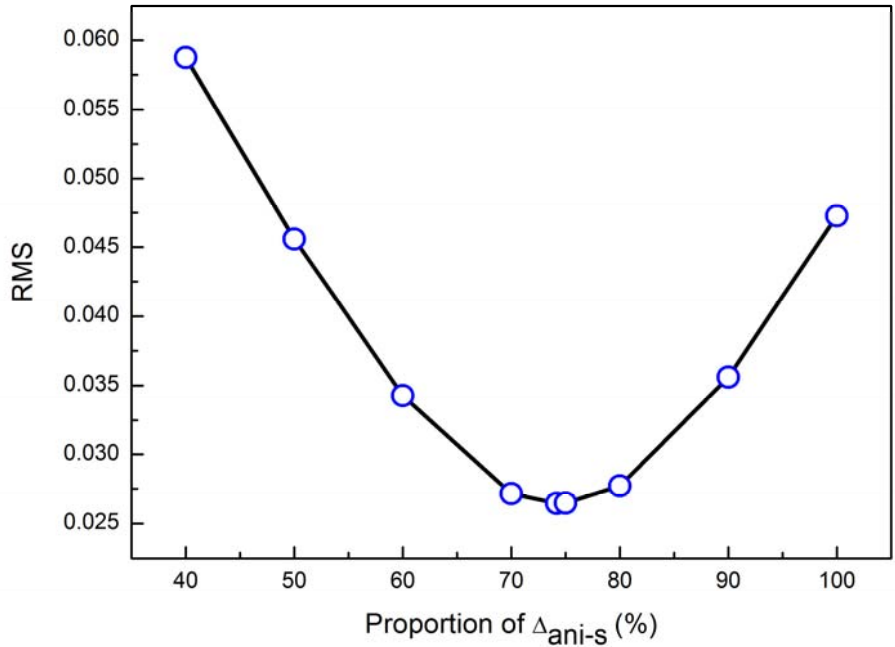

**Figure S2 | How well can the raw data be fitted by tuning the proportion of $\Delta_{ani-s}$.** The x-axis represents the proportion of the component due to the gap $\Delta_{ani-s}$. The y coordinate represents the root-mean-square (RMS) value of the difference between experimental data and the fitting results. The smaller the RMS is, the better the fitting will be. The combination of 26%$\Delta_s$+74%$\Delta_{ani-s}$ turns out to be the best fit.

**Supplementary Information III**

We fitted the $n^{th}$ LL peak energies ($E_n$) by using the equation:

$$E_n = E_D + v_F\sqrt{2eB\hbar|n|}$$

where $v_F$ is the Fermi velocity and $B$ = 5 T is the magnetic field. The LL peak energy data are obtained from STS measured at 5 T at different positions. The fitting results are displayed in Fig.S3. From the fitting, we can get the average value of $E_D$ (Dirac point energy) which is about -340 meV and $v_F$ is about $6.7 \times 10^5$ m/s. These values are qualitatively consistent with previous studies of $Sr_xBi_2Se_3$[2] and $Bi_2Se_3$[3].

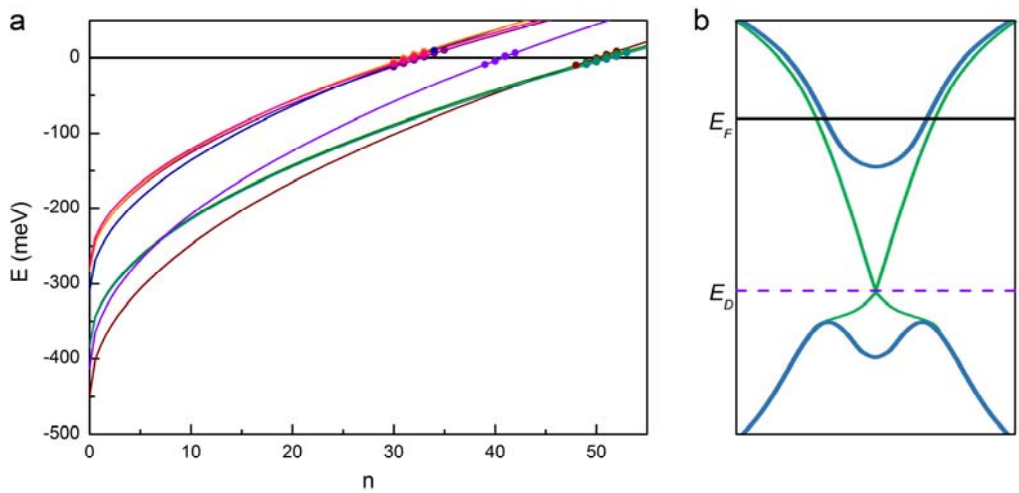

**Figure S3 | Theoretical fittings to LLs peak positions at 5T. a**, The dots represent the LL peak energies obtained from the STS measured at 5 T. the solid curves are the fitting results. The n value here represents the $n^{th}$ Fermi level. **b,** Schematic illustration of the band structure of $Sr_xBi_2Se_3$. The topological SSs are indicated by band marked with the green curves. $E_F$ and $E_D$ represent the Fermi energy and the Dirac point energy.

**Supplementary Information IV**

To visualize the effect of the LLs on the spectrum, a proper way to normalize the spectra in Fig.4 is very important. Usually, the STS measured at zero field in the same area at a high energy outside the gap is smooth and has little difference, for example, one can see the data in Fig.4a. It is easy to normalize all the curves by selecting the *dI/dV* value at a certain voltage as a reference. While because of the fluctuations due to the LLs at a finite magnetic field in the present case, this becomes nontrivial, because the curve at a high energy fluctuates a lot, which make it difficult to decide at which point all the curves should be normalized. We have tried several schemes and developed an efficient one as addressed below.

For a fixed magnetic field, we first calculate the integral of the spectrum measured at each location between -10 meV to 10 meV, then normalize these spectra measured at different spatial locations by taking the integral as the normalizing factor. Then we obtain the average spectra at each magnetic field. Finally all the spectra at the same field are normalized to unity by dividing the value of *dI/dV* on the averaged spectrum at the bias voltage of 10 meV.